\documentstyle[epsf,twoside,fleqn,espcrc2]{article}

\newcommand{\beq}{\begin{equation}}
\newcommand{\eeq}{\end{equation}}

\newcommand{\eq}{\ref}

\newcommand{\asms}{\alpha_{{}^{\overline{MS}}}}
\newcommand{\MSB}{\overline{MS}}
\newcommand{\MOM}{\widetilde{MOM}}

\newcommand{\Lam}{\Lambda_{\widetilde{MOM}}}
\newcommand{\Lams}{\Lambda_{\overline{MS}}}

\newcommand{\Gev}{{\rm GeV}}

\newcommand{\ewxy}[2]{\setlength{\epsfxsize}{#2}\epsfbox[-10 110 550
590]{#1}}
\newcommand{\AmS}{{\protect\the\textfont2
  A\kern-.1667em\lower.5ex\hbox{M}\kern-.125emS}}

\title{
Status of $\alpha_s$ determinations from the non-perturbatively
renormalised three-gluon vertex} 

\author{C. Parrinello\address{ Dept. of Mathematical Sciences, University of Liverpool, 
        Liverpool L69 3BX, United Kingdom}%
        and 
        D. G. Richards\address{Dept.\ of Physics \& Astronomy, University of Edinburgh, Edinburgh EH9 3JZ, United Kingdom} (UKQCD Collaboration), \\
B. All\'es\address{Dept. di Fisica, Sezione Teorica, Universit\`a di Milano, 
Via Celoria 16, 20133-Milano, Italy}, %
H. Panagopoulos\address{Department of Natural Sciences, University of Cyprus, CY-1678 Nicosia, Cyprus}
and C. Pittori\address{L.P.T.H.E., Universit\'e de Paris Sud, Centre d'Orsay, 91405 Orsay, France}}
       
\begin{document}

\begin{abstract}
We demonstrate the feasibility of computing $\alpha_s$
from the lattice three-gluon vertex in the Landau gauge. 
Data from $16^4$ and $24^4$ quenched lattices at $\beta=6.0$ are presented.
Our main result is that 2-loop asymptotic scaling is observed 
for momenta in the range $1.8-2.3 \ {\Gev}$, where 
lattice artifacts 
appear to be under control.
\end{abstract}

\maketitle

\section{THE METHOD}
$\alpha_{s}$ can be extracted from the 3-gluon vertex \cite{io} 
by evaluating two-point and three-point off-shell
gluon Green's functions on the lattice 
 and imposing non-perturbative renormalisation 
conditions on them, for different values of the external momenta.
The main advantages of this method are the possibility to  
 obtain $\alpha_{s} (\mu)$ 
at several momentum scales $\mu$
 from a single simulation 
and the fact that lattice perturbation theory (LPTH) is not
needed to match our coupling to $\asms$.
Also, the method can be applied with no modifications to the unquenched 
case.
On a more technical level, additional advantages are that, since one 
works in momentum space, lattice artifacts 
can be carefully analysed, as shown is section 2.1, 
and since only gluonic operators are used, the technology 
is simple.
The method is fully described in \cite{npb}. Working in the Landau gauge
and using a suitable 
definition 
 of the lattice gluon field $A_{\mu}$,
one can define the momentum space gluon 
propagator
$G^{(2)}_{U \ \mu \nu}(p) \equiv T_{\mu \nu}(p)\ G_{U} (p^2)$, where 
$T_{\mu \nu}(p)$ is the transverse projector, 
 and the complete gluon three-point function
$G_{U \ \alpha \beta \gamma}^{(3)} (p_{1}, p_{2}, p_{3})$. 
 The propagator is non-perturbatively renormalised 
by imposing that 
for $p^2=\mu^2$ it attains its
continuum tree-level value:
 \begin{equation}
G_R (p)\vert_{p^2=\mu^2}
= Z^{-1}_{A} (\mu a) G_U (p a)\vert_{p^2=\mu^2}
\nonumber \\
=\frac{1}{\mu^2}.
\label{eq:za}
\end{equation}

The three-gluon vertex is evaluated at 
the asymmetric kinematical points where, 
based on the general form of continuum the vertex function  
in the Landau gauge \cite{general}, the following relation holds:
\begin{equation}
\frac{\sum_{\alpha=1}^{4} \ G_{U \ \alpha \beta \alpha}^{(3)}
 (p a, 0, -p a)}{(G_{U} (p a))^2 \ G_{U} (0)} =
6 \ i \ Z_{V}^{-1} (pa) \ g_{0} \ p_{\beta}. 
\label{eq:baba}
\end{equation}
By computing the above ratio one can determine
 the factor  $Z_{V}^{-1} \ g_{0}  $, where $Z_V$ 
is the vertex renormalisation constant.
Finally, one can define the running coupling $g$ at the scale 
$\mu$ from the renormalised three-gluon vertex as
\beq g(\mu) = Z_A^{3/2}(\mu a) \ Z_{V}^{-1} (\mu a) \ g_{0}.
\protect\label{eq:renvert}
\eeq 
 and $\alpha_s(\mu) \equiv g(\mu)^2/4 \pi$. 
This choice corresponds to a momentum subtraction 
scheme, usually referred to as $\MOM$ in continuum QCD \cite{HH}.

\section{NUMERICAL RESULTS}

We consider two quenched data sets at $\beta = 6.0$; 150 
configurations on a $16^4$ lattice, and $103$ configurations 
on a $24^4$ lattice.  
We used a Landau gauge fixing overrelaxation algorithm, with the final iterations being
performed in double precision. It is worth emphasising that the 
numerical accuracy of the gauge-fixing is crucial to obtain a 
good signal for the three-point function. 

\subsection{Analysis of Lattice Artifacts}

\begin{table*}[htb]
\setlength{\tabcolsep}{1.0pc}
\newlength{\digitwidth} \settowidth{\digitwidth}{\rm 0}
\catcode`?=\active \def?{\kern\digitwidth}
\caption{Symmetry tests for $G^{(3)}_{U \ \alpha\beta\gamma}(p,0,-p)$ 
on the $16^4$ lattice at $\beta = 6.0$ 
(C=CPTH, L=LPTH, N=numerical).}
\protect\label{tab:o4_sym_3pt}
\begin{tabular*}{\textwidth}{@{}l@
{\extracolsep{\fill}}
rrrrrrrrr}
\hline
                 & \multicolumn{3}{c}{$G_{010}/G_{111}$}
                 & \multicolumn{3}{c}{$G_{101}/G_{111}$}
                 & \multicolumn{3}{c}{$G_{011}/G_{111}$} \\
\cline{2-4} \cline{5-7} \cline{8-10}
                 & \multicolumn{1}{r}{C}
                 & \multicolumn{1}{r}{L}
                 & \multicolumn{1}{r}{N}
                 & \multicolumn{1}{r}{C}
                 & \multicolumn{1}{r}{L}
                 & \multicolumn{1}{r}{N}
                 & \multicolumn{1}{r}{C}
                 & \multicolumn{1}{r}{L}
                 & \multicolumn{1}{r}{N}                      \\
\hline
$(1,1,0,0)$ & 1 & 1 & 1.000 & 1 & 1 & 1.1(2) & 1 & 1 & 1.000 \\
$(1,2,0,0)$ & 4 & 3.848 & 3.848 & $1/2$ & 0.510 & 0.3(1) & 2 & 1.962 &
1.962 \\
$(2,2,0,0)$ & 1 & 1 & 1.000 & 1 & 1 & 0.8(3) & 1 & 1 & 1.000 \\ \hline
\end{tabular*}
\end{table*}

 A comparison of numerical values of ratios of 
Green's functions tensor components 
with the theoretical expectations from LPTH and 
continuum perturbation theory (CPTH) provides a check for discretisation 
errors. In the case of the gluon propagator, the numerical 
values are in perfect
agreement with LPTH. 
As for the three-gluon vertex,
in Table 1 we show various
ratios of tensor components on the $16^4$
lattice, where in the first column the lattice momentum components 
are specified. Unless otherwise noted, the 
uncertainty is less than one unit in the last quoted figure.
Only the ratio $G_{101}/G_{111}$ is in poor agreement with LPTH.
This is related to the fact that the Landau gauge condition does not 
fix this ratio. Overall, the tensor structure is the one expected
from LPTH. This enables us to estimate the size of the lattice
artifacts and to identify a "continuum window" in momentum space. 


\subsection{Renormalisation Constants and Running Coupling}
\begin{figure}[t]
\vspace{2mm}
\ewxy{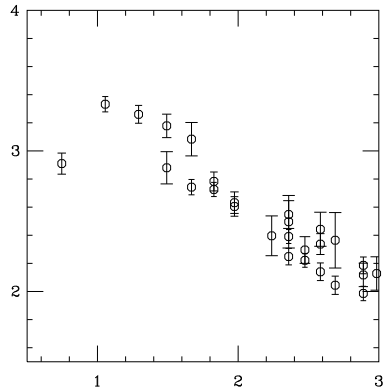}{55mm}
\caption{Running coupling $g(\mu)$ vs.\ $\mu$ on the smaller lattice.}
\label{fig:coupling}
\end{figure}

In Figure~\protect\ref{fig:coupling} we plot $g(\mu)$ on the $16^4$ lattice. 
All data are plotted vs.\ $\mu 
= \sqrt{p^2}$, expressed in $\Gev$. 
In order to detect violations of rotational invariance, we 
have used whenever possible different combinations of lattice 
momentum 
vectors for a fixed value of $p^2$, plotting separately the corresponding 
data points.
%

We obtain a clear signal, 
although the data show some
 violation of rotational invariance.

The important question is whether there is a momentum range 
where our coupling runs
according to the two-loop expression
\begin{equation}
\frac{1}{g^2(\mu)} = b_0\ln(\frac{\mu^2}{\Lam^2})+
\frac{b_1}{b_0}\ln\ln(\frac{\mu^2}{\Lam^2}),
\label{eq:twolooprun}
\end{equation}
where $b_0=11/16\pi^2$, $b_1=102/(16 \pi^2)^2$ and $\Lam$ is the QCD
scale parameter for our scheme.
To investigate this point, 
 we compute $\Lam$ as a function of the measured
values of $g^2(\mu)$ according to the formula
\begin{equation}
\Lam = \mu \ {\rm exp} \left(-\frac{1}{2 b_0 g^2(\mu)}\right)
\left[ b_0 g^2(\mu)\right]^{-\frac{b_1}{2 b_0^2}}.
\label{eq:twolooplam}
\end{equation}

If the coupling runs according to (\ref{eq:twolooprun}), then $\Lam$
as defined above approaches a constant value when
$\mu \rightarrow \infty$.
By plotting $\Lam$ versus $\mu$ (see Figure~2), 
one notices that 

\begin{figure}[htb]
\vspace{2mm}
\ewxy{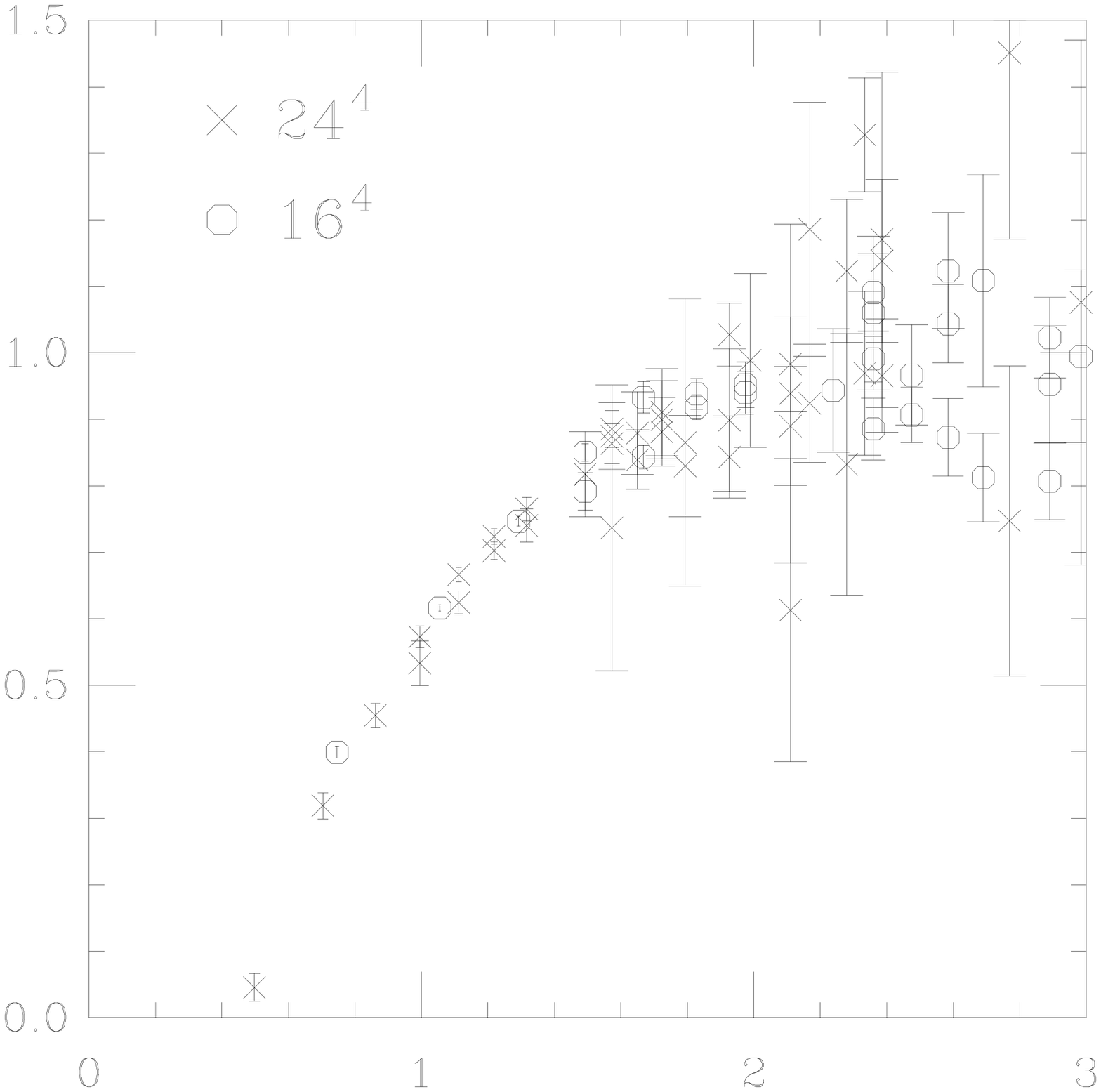}{55mm} 
\caption{$\Lam$ vs.\ $\mu$ for our two lattice sizes.}
\protect\label{fig:lambda}
\end{figure}
for $\mu < 1.8 \ \Gev$, $\Lam$ depends strongly on $\mu$. 
However, in the range $1.8 < \mu < 2.3 \ \Gev$ the data are consistent with a
constant value for $\Lam$. 
No violations of rotational invariance are
observed and a comparison of the two lattice sizes shows
no volume dependence. 
 For $\mu > 2.3 \ \Gev$, rotational invariance is broken by higher
order terms in $a^2$ and the two-loop behaviour disappears.
In summary, we appear to have a ``continuum window'' in the range
$1.8 < \mu < 2.3  \ \Gev$, where two-loop
scaling is observed and lattice artifacts are under control.
In order to extract a prediction for $\Lam$, we fit the data points in
the continuum window to the curve obtained by inserting (\ref{eq:twolooprun}) 
in (\ref{eq:twolooplam}).
We take as our best estimate
the fit to the $16^4$ data, for which the statistical errors are
smaller, and obtain
\beq
\Lam=0.88 \pm 0.02
\pm 0.09 \ \Gev,
\label{eq:resu}
\eeq
where the first error is statistical and the second error comes from the
uncertainty on the value of $a^{-1}$.

\section{MATCHING TO $\overline{MS}$}
\label{section:matching}
\par
We can extract a prediction for $\asms$ with zero quark
flavours from our numerical
results for $\Lam$.
 The ratio $\Lams/\Lam$
 can be determined to all orders in the coupling
constant from a one-
loop continuum calculation.
We obtain, in the Landau gauge and for zero quark flavours
$ \Lams^{(0)}/\Lam^{(0)}=0.35$.
This result  is in agreement with
 previous one-loop calculations of the three-gluon vertex \cite{BF}.
Using (\eq{eq:resu}) and the above result, we obtain 
\beq
\Lams^{(0)}= 0.31 \pm 0.05 ~\Gev.
\eeq
This is the main result of our computation, which is in very good 
agreement  with 
the one of ref. \cite{Bali}.
In terms of $\asms^{(0)}$, our result yields:
\begin{equation}
\asms^{(0)}(2.0 ~\Gev)= 0.24 \pm 0.02.
\end{equation}

Although LPTH is not needed for matching, we can use it to perform 
some interesting cross-checks of the continuum calculation 
(see \cite{npb} for details).

\section{CONCLUSIONS}
\label{section:conclusions}
We have shown in the quenched approximation that a non-perturbative
determination of the QCD running coupling can be obtained from first
principles by a lattice study of the three-gluon vertex.  We have
some evidence that systematic lattice effects are under control in our
calculation. LPTH is not
needed to match our results to $\MSB$ and the extension to the
full theory does not present in principle any extra problem.  

C. Parrinello 
and DGR acknowledge support from PPARC through travel grant GR/L29927 and 
Advanced Fellowships. We also acknowledge support from EPSRC grants GR/K41663 
and GR/K55745.

\end{document}